\documentclass[]{pasj02} 
\usepackage[switch,mathlines]{lineno} 
\usepackage{bm,url,color}
\usepackage{subcaption}
\usepackage{xcolor}

\jyear{2025}
\Received{}
\Accepted{}

\begin{document} 

\title{Galaxy fly-bys sustain bar–halo friction and bar slowdown in disk galaxies}

\author{
Rumi \textsc{Kodama},\altaffilmark{1}
Rimpei \textsc{Chiba},\altaffilmark{1}\altemailmark \orcid{0000-0002-3445-855X}\email{rimpei-chiba@g.ecc.u-tokyo.ac.jp}
Tetsuro \textsc{Asano},\altaffilmark{2,3,4}\altemailmark \orcid{0000-0002-7523-064X}\email{asano@fqa.ub.edu}
Junichi \textsc{Baba},\altaffilmark{5, 6}\altemailmark \orcid{0000-0002-2154-8740}\email{junichi.baba@sci.kagoshima-u.ac.jp}
 and 
Michiko S \textsc{Fujii}\altaffilmark{1}\altemailmark\orcid{0000-0002-6465-2978} \email{fujii@astron.s.u-tokyo.ac.jp}
}
\altaffiltext{1}{Department of Astronomy, Graduate School of Science, The University of Tokyo, 7-3-1 Hongo, Bunkyo-ku, Tokyo 113-0033, Japan}
\altaffiltext{2}{Departament de F\'isica Qu\`antica i Astrof\'isica (FQA), Universitat de Barcelona (UB), c. Mart\'i i Franqu\`es, 1, 08028   Barcelona, Spain}
\altaffiltext{3}{Institut de Ci\`encies del Cosmos (ICCUB), Universitat de Barcelona (UB), c. Mart\'i i Franqu\`es, 1, 08028 Barcelona, Spain}
\altaffiltext{4}{Institut d’Estudis Espacials de Catalunya (IEEC), c. Gran Capit\`a, 2-4, 08034 Barcelona, Spain}
\altaffiltext{5}{Amanogawa Galaxy Astronomy Research Center (AGARC), Kagoshima University, 1-21-35 Korimoto, Kagoshima 890-0065, Japan}
\altaffiltext{6}{National Astronomical Observatory of Japan, Mitaka-shi, Tokyo 181-8588, Japan}





\maketitle

\begin{abstract}
Bars in disk galaxies slow down as they transfer their angular momentum to their dark matter halo via dynamical friction from near-resonant orbits.
This bar-halo dynamical friction can become ineffective once phase mixing erases the phase-space gradient around the main resonances.
We present fully self-consistent $N$-body simulations of a Milky Way-like disk galaxy with a single dwarf-galaxy fly-by in prograde and retrograde orbits before, during, and after bar formation. 
In our models, the fly-bys do not trigger a long-lived tidal bar; the bar forms on essentially the same time as in the isolated model. After the encounter, however, all perturbed models develop bars that are stronger and slower than in the isolated one. 
The final pattern speed depends little on the encounter time, but it does depend on the encounter direction relative to the disk rotation: prograde encounters slow the bar more than retrograde ones. The angular-momentum evolution shows that the disk loses its angular momentum and the halo gains it, consistent with bar-halo friction.
By probing the particle distribution of the halo in angle-action space, we demonstrate that the isolated bar enters a metastable, saturated state with a flattened distribution in the phase space around the bar's corotation resonance, whereas a dwarf passage excites long-lived fluctuations in the halo that restore the phase-space gradients near the corotation and thereby sustain the bar-halo friction. 
This mechanism explains the continued slowdown and growth of bars after fly-bys. It may be relevant to the Milky Way, whose bar formed near the epoch of a major ancient accretion event, suggesting that an early encounter could have influenced the subsequent secular evolution of the bar.
\end{abstract}


\section{Introduction}\label{sec:introduction}
Bars are common in disk galaxies and are key drivers of their secular evolution \citep{2004ARA&A..42..603K}. 
They torque the gas, removing its angular momentum and funneling it toward the central kiloparsec, where it fuels nuclear star formation (e.g., \cite{1992MNRAS.259..345A,2020MNRAS.497.5024S,2020MNRAS.492.4500B}). 
Bars also redistribute the angular momentum of both stars and gas across the whole disk, transferring it between the inner and outer regions and thereby reshaping the global mass and kinematic structure and reorganizing the gas distribution (e.g., \cite{1995A&A...301..649F}). This secular rearrangement can drive strong radial migration of stars and gas (\cite{2013A&A...553A.102D,2025PASJ...77..916B,2025arXiv250819340B}) and suppress star formation in parts of the inner disk (\cite{
2017MNRAS.465.3729S,2022MNRAS.513.2850B}).

Bars in isolated galaxies arise from the disk's gravitational instability \citep{OP1973,efstathiou1982stability} and then evolve secularly as they transfer angular momentum to the outer disk and, crucially, to the live dark matter halo by dynamical friction \citep{athanassoula2002bar,2014RvMP...86....1S}. Dynamical friction is essentially a resonant phenomenon in which a net torque arises from the imbalance between the number of near-resonant particles that gain and lose angular momentum as they librate around the resonances (e.g., \cite{lynden1972generating,Tremaine1984Dynamical,weinberg1985evolution,Chiba2023}). The efficiency of this angular momentum transfer hence depends on the gradient of the phase-space density at the main resonances, which is set by a variety of factors, including the halo's central density profile \citep{2000ApJ...543..704D,2002MNRAS.330...35A}, its velocity anisotropy \citep{Sellwood2016BarInstability}, and its net rotation \citep{2013MNRAS.430.2039S,2014ApJ...783L..18L,fujii2018,ChibaKataria2024}.

Since the phase-space density of systems in self-gravitational equilibrium typically has negative gradients in angular momentum (e.g., \cite{Tremaine1984Dynamical}), the particles that gain angular momentum initially outnumber those that lose it, leading to a net transfer of angular momentum from the bar to the halo. However, after a few libration times, the distribution of particles near resonances phase mixes and becomes flat. Once this occurs, the net angular momentum transfer becomes zero \citep{Chiba2022Oscillating,Banik2022Nonperturbative}, and dynamical friction is said to be ``saturated'' \citep{Goldreich1980Disk}. This nonlinear saturation of bar-halo friction has been reported in a number of earlier simulations (e.g., \cite{Sellwood2006Metastability,Beane2023StellarBars}).

Real galactic disks are, however, not isolated.
They are surrounded by satellite galaxies and thus regularly experience encounters and associated tidal perturbations (e.g., \cite{2017ApJ...847....4G,2024ApJ...976..117M}).
The Milky Way (MW) galaxy, which is also a barred galaxy, is a clear case: \textit{Gaia} has revealed vertical phase-space spirals, bending/breathing waves, and signatures of the reflex motion of the inner Galaxy, all of which are possibly caused by satellite-driven perturbations, in particular those by the Sagittarius dwarf galaxy and the Large Magellanic Cloud
(e.g., \cite{antoja2018dynamically,Laporte2018,2023Galax..11...59V,2025A&A...700A.109A}). 

In this paper, we investigate how perturbations by single galaxy encounters (fly-bys) affect the long-term evolution of galactic bars.
Although there are several studies investigating the role of external satellite perturbations in directly triggering bar formation or causing bar disruption (e.g., \cite{1987MNRAS.228..635N,1990A&A...230...37G,Lang2014,2017A&A...604A..75M,2017MNRAS.464.1502M,2018MNRAS.474.5645P,Lokas2018, 2025A&A...698L...7J,Chen+2026arXiv260219995C}), little is known about their indirect impact on the bar's post-formation evolution through the changes they produce in the dark halo’s phase-space distribution.
The notable exception is the work by \citet{Sellwood2006Metastability}, who showed that when the bar-halo friction saturates, a modest perturbation can disrupt the resonance and allow dynamical friction to resume. Because this saturated state was fragile to external perturbations, they referred to it as a ``metastable'' state. This phenomenon was recently studied in a general framework by \citet{Hamilton2023BarResonanceWithDiffusion}, who showed that, in noisy environments where particles undergo diffusion in phase space, the distribution near resonances never fully phase-mixes. As a result, the phase-space gradient is retained, and thus dynamical friction is kept active. On the other hand, \citet{Beane2023StellarBars} recently showed that the presence of gas, which yields angular momentum to the bar, can prevent the resonance from moving and thus help the bar-halo system settle into a metastable state, a process they coined ``bar-locking''. The goal of this paper is to investigate how perturbations induced in the halo by dwarf fly-bys may affect the bar's subsequent evolution.

The impact of dwarf fly-bys on bar evolution may be investigated using cosmological simulations, which naturally include perturbations from dwarf/satellite galaxies (e.g. \cite{2018MNRAS.473.2608Z,2018MNRAS.479.5214Z}). However, many processes, such as mergers, gas accretion, and stellar feedback, act together, making it difficult to disentangle the pure dynamical effect. In addition, resolving the detailed phase-space structure of the dark matter halo requires a large number of particles \citep{Weinberg2007}, which remains computationally too expensive for cosmological simulations. We therefore explore the dynamical impact of dwarf-galaxy fly-bys on the bar's secular evolution using a series of controlled, fully self-consistent high-resolution $N$-body simulations of a Milky Way-like disk galaxy in a live halo perturbed by a dwarf galaxy. In particular, we introduce the dwarf on either a prograde or a retrograde orbit at different evolutionary stages of the bar (before, during, and after bar formation) and compare the results with those of an isolated galaxy simulation.

Section~\ref{sec:method} describes the numerical set-up and initial conditions. Section~\ref{sec:results} presents the results: Section~\ref{sec:barevol} quantifies the bar's time evolution under perturbations, Section~\ref{sec:amtransfer} examines the encounter dependence of disk-halo angular-momentum exchange, Section~\ref{sec:dynfric} develops the resonant-torque picture and demonstrates saturation and its recovery, and Section~\ref{sec:halowake} analyzes the halo's response to both the dwarf and the bar using angle-action variables and demonstrates the saturation and recovery of dynamical friction. Finally, Section~\ref{sec:summary} summarizes our findings and discusses their implications.

\section{Methods}
\label{sec:method}

\subsection{Initial conditions}

To isolate the purely gravitational impact of dwarf-galaxy fly-bys on bar evolution, we performed collisionless $N$-body simulations of a live-halo disk galaxy with and without a passing dwarf perturber. 
Because we focus on the effect of a single perturbation caused by a dwarf galaxy on a bar rather than tidally forced bar formation, we set up a disk-galaxy model that forms a bar in isolation. 
We therefore adopted a Milky-Way galaxy model used in \citet{fujii2019modelling} as the `MW model.' 
The stellar disk follows a radially exponential, vertically isothermal profile:
\begin{align}
  \rho_{\mathrm{d}}(R, z) = \rho_{\mathrm{d}0}
  \exp \left(-\frac{R}{R_{\mathrm{d}}} \right)
  \mathrm{sech}^2 \left(\frac{z}{z_{\mathrm{d}}} \right),
\end{align}
with $R_{\mathrm{d}}=2.3\,\mathrm{kpc}$ and $z_{\mathrm{d}}=0.2\,\mathrm{kpc}$, and the disk total mass $3.9\times10^{10}\,M_{\odot}$.
The bulge follows a Hernquist profile \citep{1990ApJ...356..359H} with scale radius of $0.75\,\mathrm{kpc}$ and characteristic velocity dispersion of $330\,\mathrm{km\,s^{-1}}$, for a total bulge mass of $5.4\times10^{9}\,M_{\odot}$. 
The dark matter halo is modeled as a live $N$-body system with a Navarro-Frenk-White profile (NFW; \cite{1997ApJ...490..493N}) with scale radius of $10\,\mathrm{kpc}$, and characteristic velocity dispersion of $420~\mathrm{km\,s^{-1}}$, yielding a total halo mass of $8.7\times10^{11}\,M_{\odot}$.
To keep the bar sufficiently short in the isolated reference run, we introduced net prograde halo spin \citep{fujii2019modelling}; the spin parameter $\alpha_{\mathrm{h}}=0.8$, defined as the fraction of halo particles with $L_z>0$ (thus, $\alpha_{\mathrm{h}}=0.5$ is non-spinning).

\begin{figure*}
    \centering
    \begin{minipage}[b]{0.42\textwidth}
        \centering
        \includegraphics[width=\textwidth]{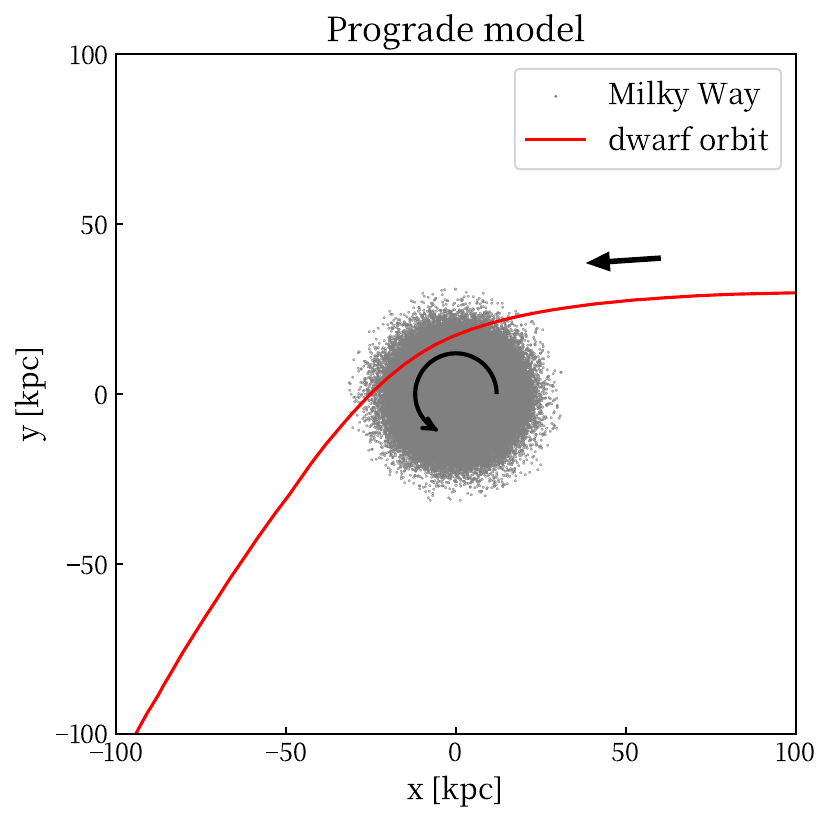}
    \end{minipage}
    \hspace{0.01\textwidth}  
    \begin{minipage}[b]{0.42\textwidth}
        \centering
        \includegraphics[width=\textwidth]{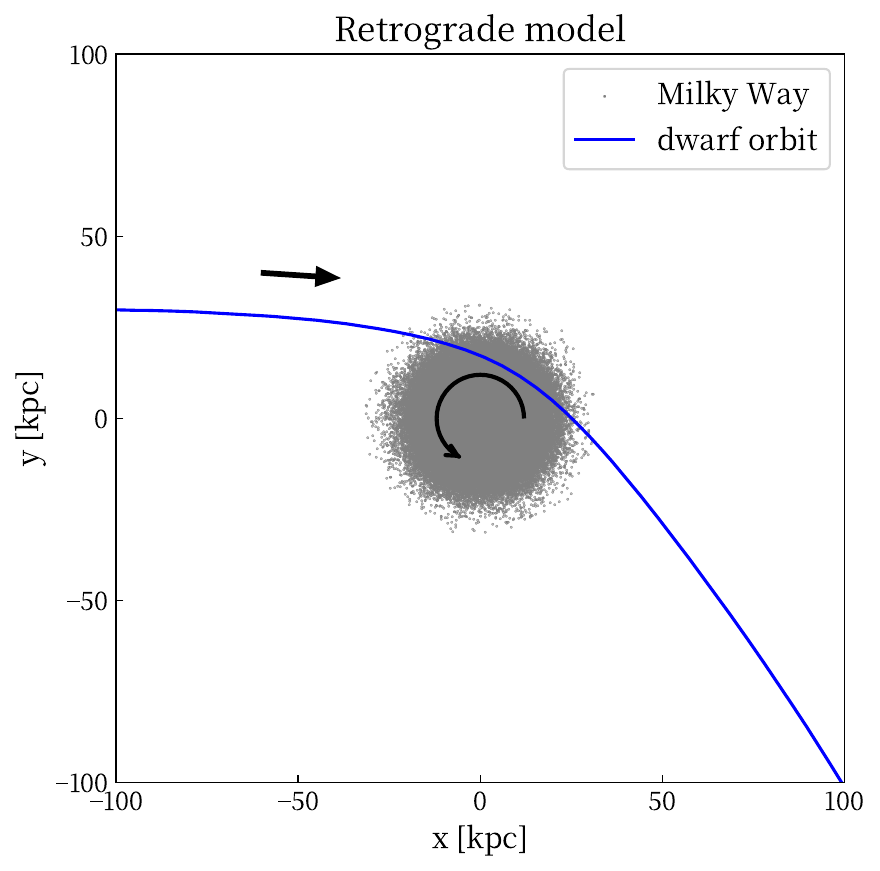}
    \end{minipage}
    \caption{The trajectory of the dwarf galaxy in prograde (left) and retrograde (right) orbits. Gray dots indicate the disk stars, and the arrow in the disk shows the rotation of the disk. The arrows above the curve indicate the direction of the motion. {\textbf{Alt text}: scatter plots showing the distribution of disk stars, lines showing the orbit of the dwarf galaxy, and arrows showing the disk rotation directions.}}
    \label{fig:orbits}
\end{figure*}

The perturber, a self-gravitating dwarf galaxy, was represented by a live NFW halo with a scale radius of $5$ kpc and total mass of $4.1\times10^{10}\,M_\odot$. We considered coplanar fly-bys in two orbital senses: prograde and retrograde with respect to the disk rotation.
For prograde encounters, the dwarf added at $(x, y, z)=(120,30,0)$\,kpc with velocity $(v_x, v_y, v_z)=(-30,0,0)$\,km\,s$^{-1}$; for retrograde encounters, at $(x, y, z)=(-120,30,0)$\,kpc with $(v_x, v_y, v_z)=(30,0,0)$\,km\,s$^{-1}$.
We added the dwarf to the isolated run at $t=0.0$, $1.0$, and $2.5$ Gyr. 
These choices yield the first pericenter passages at $t\simeq0.32$, $1.3$, and $2.8$ Gyr (labels 00, 10, 25, respectively), measured on the isolated model's clock; we denote the resulting runs p00/p10/p25 (prograde) and r00/r10/r25 (retrograde). Orbits in the disk frame are shown in Fig.~\ref{fig:orbits}.

Equilibrium initial conditions for both the MW and dwarf models were generated with \textsc{GalactICs} \citep{1995MNRAS.277.1341K,2005ApJ...631..838W,2008ApJ...679.1239W}.
Within the MW model we adopted a uniform particle mass of $\sim10^{4}\,M_\odot$ across halo, disk, and bulge, corresponding to $97$\,M, $3.9$\,M, and $0.56$\,M particles, respectively; this exceeds the $\gtrsim10^6$ disk-particle requirement for $\sim10$ Gyr integrations \citep{fujii2011dynamics}.
The dwarf model is modeled with a coarser mass resolution than the MW model because we do not have to resolve its internal evolution. 
The parameters for both galaxies are also summarized in Table~\ref{tab:models}.

\begin{table*}
    \begin{center}
	\caption{Model parameters of galaxies. From left, halo scale radius ($a_{\rm h}$), characteristic velocity dispersion ($\sigma_{\rm h}$), spin parameter ($\alpha_{\rm h}$), total mass ($M_{\rm h}$), number of particles ($N_{\rm h}$), disk scale radius ($R_{\rm d}$), scale height ($h_{\rm d}$), total mass ($M_{\rm d}$), number of particles ($N_{\rm d}$), bulge scale radius ($a_{\rm b}$), characteristic velocity dispersion ($\sigma_{\rm b}$), total mass ($M_{\rm b}$), and number of particles ($N_{\rm b}$). }
	\label{tab:models}
	\begin{tabular}{l|ccccc|cccc|cccc} 
		\hline
        &\multicolumn{5}{|l}{Halo} & \multicolumn{4}{|l}{Disk} & \multicolumn{4}{|l}{Bulge} \\
        \hline
		&$a_\text{h}$ & $\sigma_\text{h}$ &$\alpha_{\rm h}$& $M_{\rm h}$ & $N_{\rm h}$ & $R_{\rm d}$ & $h_{\rm d}$ & $M_{\rm d}$ & $N_{\rm d}$ & $a_\text{b}$ & $\sigma_\text{b}$ & $M_{\rm b}$ & $N_{\rm b}$\\
       &(kpc) & (km/s) & & $(\rm M_{\odot}$) & & 
       (kpc) & (kpc) & ($M_{\odot}$) & & 
       (kpc) & (km/s) & $(\rm M_{\odot}$) & \\
		\hline
		MW & 10 & 420 & 0.8 & $8.7\times10^{11}$ & 96M & 
        2.3 & 0.2 & $3.7\times10^{10}$ & 3.9M &
        0.75 & 330 & $5.4\times10^9$  & 0.56M \\
        Dwarf &  5  & 600 & 0.5 & $4.1\times10^{10}$  & 0.24M & - & - & - & - & - & - & -& - \\
		\hline
	\end{tabular}
    \end{center}
\end{table*}

\begin{table*}
 \begin{center}
   \caption{Summary of our models. From left, orbit of the dwarf galaxy, time to start the simulation with the dwarf, time of the pericenter passage from the beginning of the run with the dwarf, and pericenter distance of the dwarf.}
   \label{tab:6dwarf}  
  \begin{tabular}{lcccc} \hline
    Name & Dwarf orbit & Starting time & Pericenter passage & Pericenter distance \\
         &    & (Gyr)                 &      (Gyr)   &(kpc)   \\  \hline
    isolated & -  &   -                  &      -      & -\\
    p00 &  prograde   &   0                  &      0.32      & 15.8\\
    p10 &  prograde    &   1.0                 &      1.3      & 15.7\\
    p25 &  prograde   &   2.5                 &      2.8      & 15.9\\
    r00 & retrograde  &   0                  &      0.32      & 15.7\\
    r10 & retrograde  &   1.0                 &      1.3      & 15.6\\
    r25 & retrograde  &   2.5                 &      2.8      & 15.7\\  \hline
  \end{tabular}
 \end{center}
\end{table*}

\subsection{$N$-body simulations}

We performed all simulations with \textsc{Bonsai}\footnote{\url{https://github.com/treecode/Bonsai}}, a GPU-parallel tree code \citep{bonsai}, on the Pegasus system at the Center for Computational Sciences, University of Tsukuba.
For the tree code, we adopted a gravitational softening length of $0.01$\,kpc, an opening angle of $0.4$, and a shared integration time-step of $0.61$ Myr.
Snapshots were stored every 9.78~Myr.

As a reference, we first evolved the isolated disk (hereafter isolated) for 10\,Gyr.
To examine timing effects, we restarted from isolated snapshots at $t=0$, $1.0$, and $2.5$\,Gyr, added the dwarf galaxy.
These start times correspond, respectively, to epochs before, during, and after bar formation in the isolated case.
All the simulations were performed up to 10\,Gyr from the initial time of the isolated model.

\section{Results}
\label{sec:results}

\subsection{Time evolution of the bar under perturbations}
\label{sec:barevol}

\begin{figure*}
	\includegraphics[width=\textwidth]{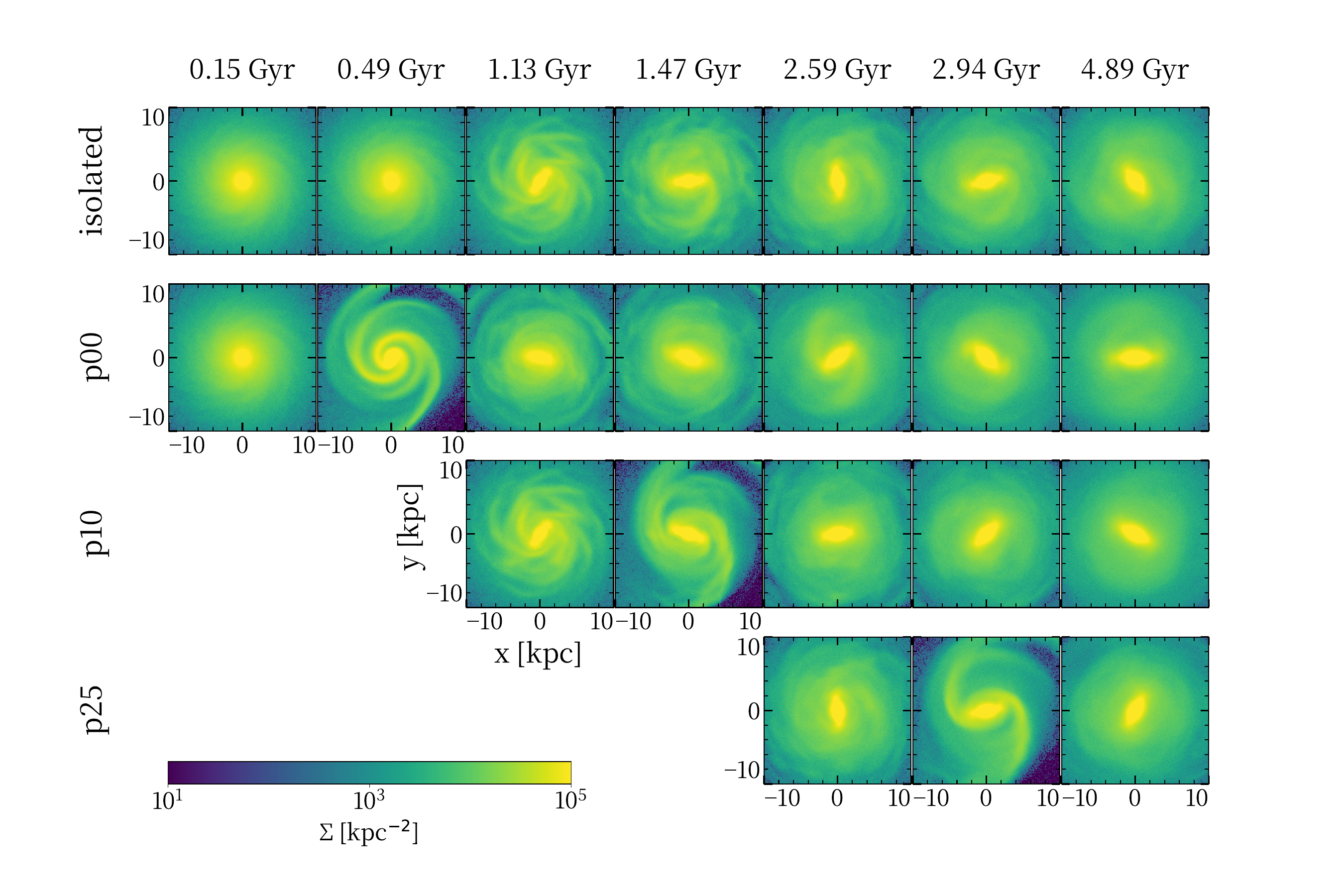}
    \includegraphics[width=\textwidth]{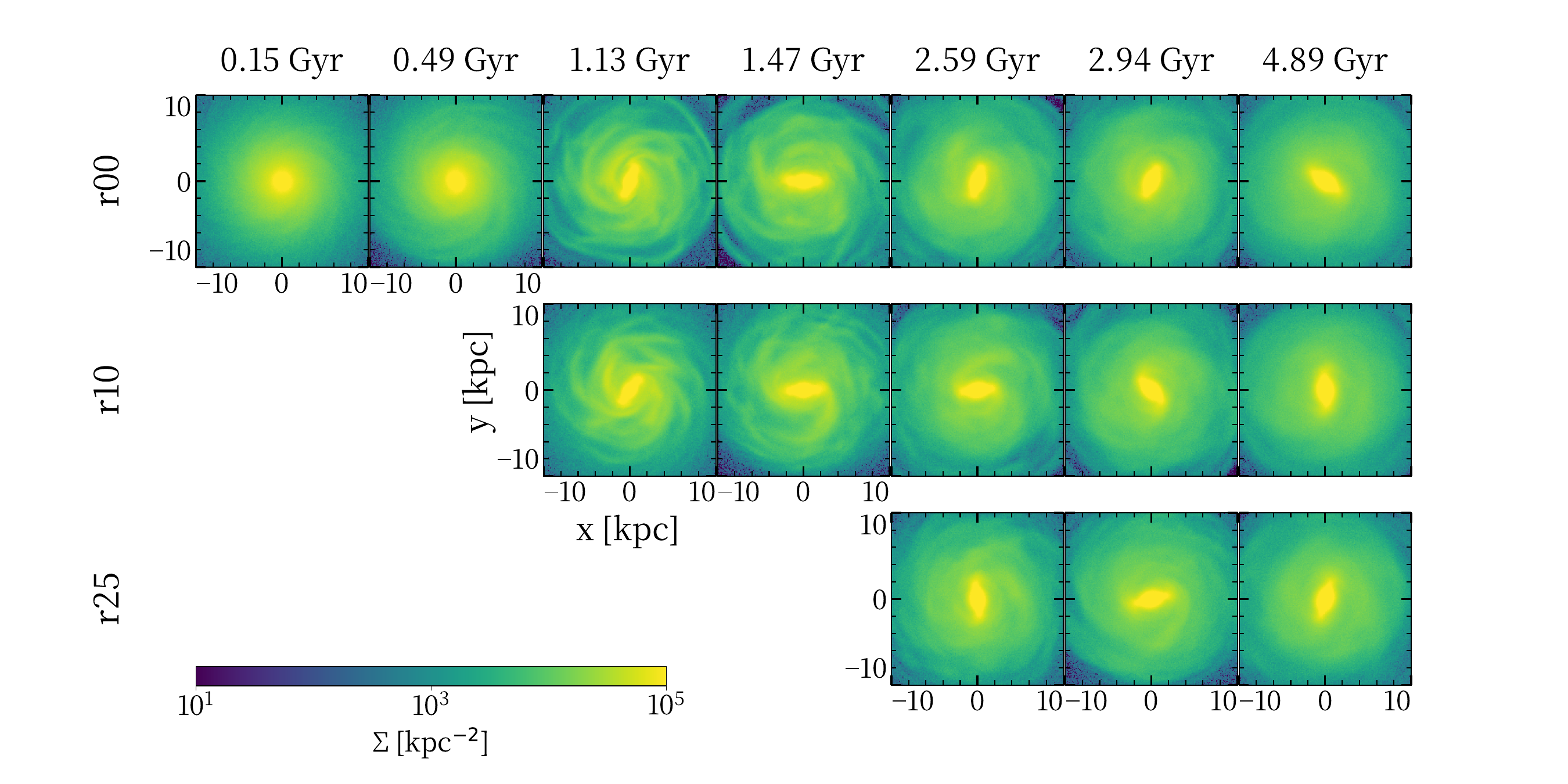}
    \caption{Snapshots of the discs for models isolated, p00, p10, p25, r00, r10, and r25 from top to bottom. The time indicates the time from the beginning of the isolated model simulation. The second left panel shows the snapshot just after the pericentre passage (0.32, 1.3, and 2.8\,Gyr for models 00, 10, and 20) of the dwarf. {\textbf{Alt text}: surface density plots for 0.15, 0.49, 1.13, 2.59, 2.94, 4.89, and 8.81 Gyr.}}
    \label{fig:snap}
\end{figure*}

Figure~\ref{fig:snap} shows representative snapshots of the simulations. 
In the isolated model, a bar appeared at $t\sim1$ Gyr and grows thereafter. 
In the prograde models, the dwarf passed pericenter at $t\approx0.32$ (p00), $1.3$ (p10), and $2.8$ Gyr (p25), and each passage excited a prominent two-armed tidal spiral that fades with time. 
By contrast, the retrograde models (r00, r10, r25) showed much weaker spirals, because the perturber's large relative angular speed makes the force non-resonant, so the torque largely averages out (e.g., \cite{1972ApJ...178..623T,1990A&A...230...37G}).
Because our MW model is intrinsically bar-unstable, a bar formed even without any fly-by encounter. 
The prograde fly-bys mainly excited short-lived tidal spirals and did not hasten bar formation; even when a passage occurred before bar formation (p00), the bar appeared at essentially the same time as in the isolated case.

To quantify the bar's amplitude and pattern speed, we decomposed the stellar surface density into a Fourier series:
\begin{equation}
  \Sigma(R,\phi)=\sum_{m=0}^{\infty} A_m(R) \exp \{i m[\phi-\phi_m(R)]\},
  \label{eq:Fourier_Decomp}
\end{equation}
where $A_m(R)$ and $\phi_m(R)$ are the amplitude and phase of the $m$th mode. 
We associated the inner $m=2$ mode with the bar and defined the bar angle as the phase $\phi_2$ averaged over $R<3.5$ kpc at each snapshot. 
The pattern speed $\Omega_{\rm p}$ is then obtained by taking the finite difference of this angle between successive snapshots.

\begin{figure}
	\includegraphics[width=\columnwidth]{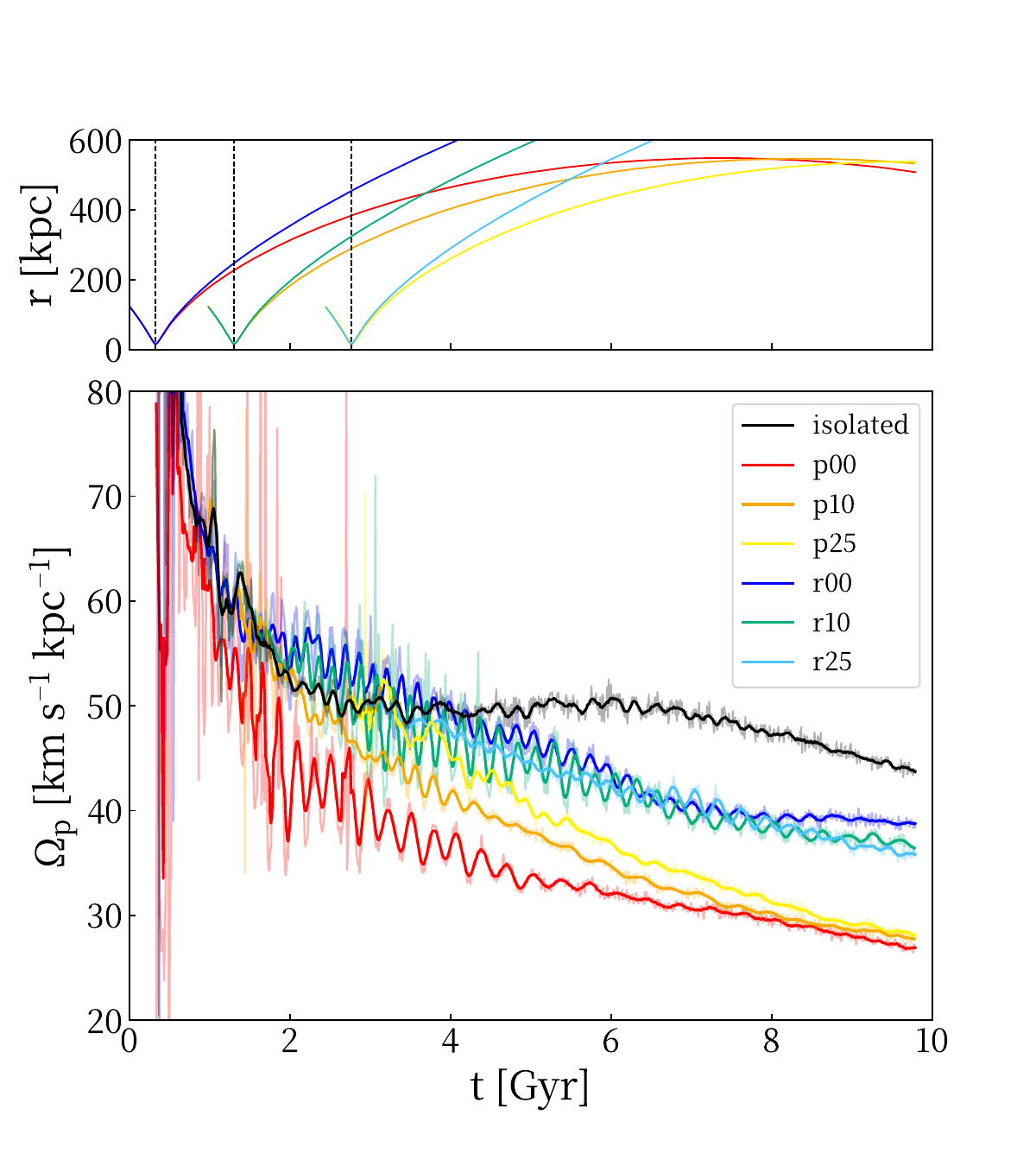}
    \caption{
    Top: The distance between the dwarf and disk center as a function of time. 
    Bottom: Time evolution of the bar pattern speed. The colors indicate each model; black, red, orange, yellow, blue, green, and cyan correspond to models isolated, p00, p10, p25, r00, r10, r25, respectively. Vertical dotted lines mark the pericenter-passage times (0.32, 1.3, and 2.8 Gyr, respectively) for the $t=0$, 1.0, and 2.5 Gyr starting-time models. {\textbf{Alt text}: The top panel is a six-line graph. The bottom panel is a line graph with seven lines showing the pattern speeds of the bars of the seven models as a function of time, 0 to 10 Gyr. The pattern speeds slow down with time, but the slowdown stalls in the isolated model.}
    }
    \label{fig:pattern_speed}
\end{figure}

Figure~\ref{fig:pattern_speed} compares $\Omega_{\rm p}$ across all models.
The thin lines show the raw data, and the thick lines show the values averaged over $\pm 50$ Myrs.
The upper panel gives the dwarf-host separation for reference, and vertical dotted lines in both panels mark the pericenter passage times (0.32, 1.3, and 2.8 Gyr, respectively). 
At early times ($t\lesssim 1$ Gyr), the pattern speed was noisy because the bar was not yet established and the $m=2$ phase was poorly defined. In the isolated model, $\Omega_{\rm p}$ initially decreased, but then stayed nearly constant at $\sim 50$\,km\,s$^{-1}$\,kpc$^{-1}$ for a few Gyrs before slowing down again after $\sim 7$\,Gyr. 
In contrast, all other perturbed runs avoided this plateau; $\Omega_{\rm p}$ continued to decrease and ended below the isolated value. 
There is a clear dependence of the final $\Omega_{\rm p}$ on the dwarf's orientation: the prograde fly-bys always yielded the lowest final $\Omega_{\rm p}$, while retrograde ones ended at intermediate values.
However, the final $\Omega_{\rm p}$ had little dependence on the dwarf's encounter time: runs with the same orbital orientation converged to similar endpoints.

\begin{figure}
	\includegraphics[width=\columnwidth]{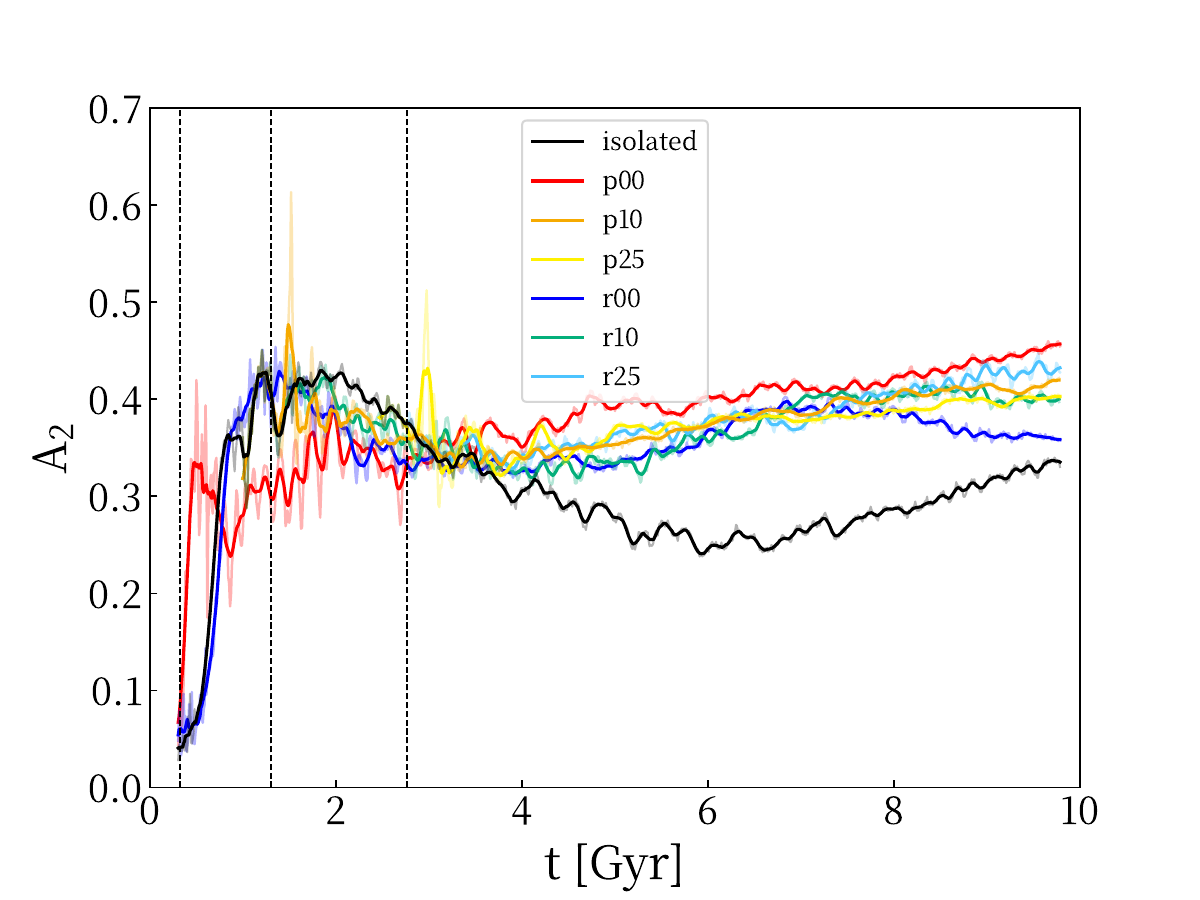}
    \caption{
    Time evolution of the $m=2$ mode maximum amplitude, $A_{2,\rm max}$. Colors and vertical lines are the same as Figure~\ref{fig:pattern_speed}.
    {\textbf{Alt text}: Graph with seven lines showing the maximum Fourier amplitude for $m=2$ as a function of time between 0 and 10 Gyr. Perturbed models show a Fourier amplitude higher than that of isolated model.}}
    \label{fig:amplitude}
\end{figure}

Figure~\ref{fig:amplitude} shows the corresponding time evolution of the bar amplitude, $A_{\rm 2,max}$, defined as the peak of $A_{\rm 2}(R)$. In the isolated model, $A_{\rm 2,max}$ rose from $t\!\sim\!1$~Gyr, reached a first maximum near $t\!\sim\!2$~Gyr, declined while $\Omega_{\rm p}$ is near-constant, and then exhibited modest regrowth as the bar slowed again. Unlike the isolated run, the perturbed bars did not undergo a pronounced mid-time decline; instead, their amplitude gradually increased, regardless of the orientation or time of the dwarf encounter.
In the prograde models (p00, p10, and p25), sharp, transient spikes appeared just after the pericenter passage, marked by the dotted vertical lines, which are signatures of the two-armed tidally-induced spirals noted above.
In contrast, the retrograde models, which exhibited weaker tidal spirals, showed no such peaks.

\subsection{Time evolution of the disk's angular momentum}
\label{sec:amtransfer}

\begin{figure}
	\includegraphics[width=\columnwidth]{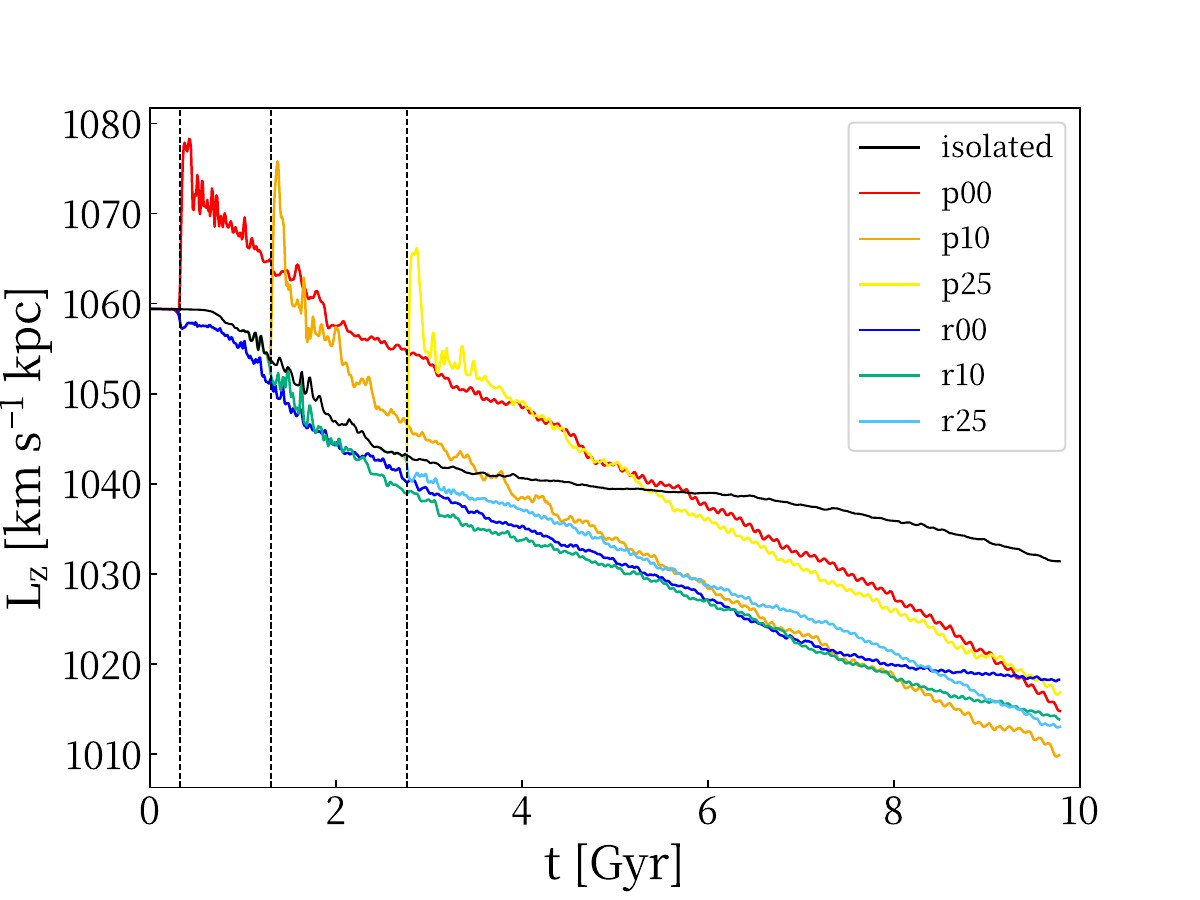}
    \caption{Time evolution of the $z$-component of the disk angular momentum. Colors and vertical lines are the same as Figure~\ref{fig:pattern_speed}. {\textbf{Alt text}: Line graph with seven lines showing the z-component of the disk angular momentum for the isolated and perturbed models as a function of time for 0 and 10 Gyr. The angular momentum transfer stalled in the isolated model but continued in the perturbed models.}}
    \label{fig:disc_Lz}
\end{figure}

To examine whether the differences in the bar evolution reflect changes in bar-halo friction, we quantified the angular momentum transfer from the bar to the dark matter halo by tracking the specific $z$-angular momentum, $L_{z}$, of the stellar disk.
Figure~\ref{fig:disc_Lz} shows the results. 
In the isolated model, $L_z$ dropped rapidly from the bar formation time ($t\approx 1$ Gyr) to $\approx 3$ Gyr, after which the decline became small, and the curve was nearly flat. 

In the prograde models, each pericenter passage caused a clear upward jump in $L_z$, arising from the dwarf's direct tidal torque. Immediately after the passage, $L_z$ was higher than that of the isolated model, but it continued to decline and ended below the isolated value.
In the retrograde models, the direct impact of the dwarf caused a much smaller downward dip in $L_z$.
Similar to the prograde cases, the disk's $L_z$ continued to decline without forming a plateau.
Despite the distinct response against the dwarf passage, both prograde and retrograde models converged to nearly the same $L_z$ with little dependence on the encounter time. However, the converged pattern speed of prograde ones was slower than that of retrograde ones.  

These results imply that the direct transfer of angular momentum from the dwarf to the disks occurred on a very short timescale and had little impact on the disks' final angular momentum. However, the final angular momentum was always smaller than that of the isolated model, which suggests that the dwarf fly-by can have a long-lasting, indirect impact on the bar’s secular evolution by affecting the efficiency of bar-halo interaction.

\subsection{Saturation and recovery of dynamical friction}
\label{sec:dynfric}

\begin{figure*}
    \centering
      \begin{subfigure}{0.33\textwidth}
        \caption*{\large Initial ($t = 0~\mathrm{Gyr}$)}
        \includegraphics[width=\textwidth]{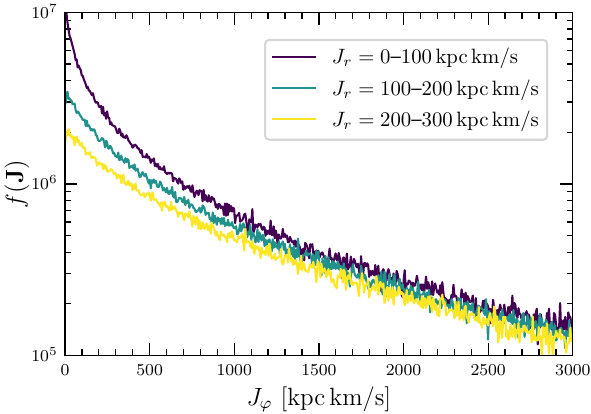}
      \end{subfigure}
      \begin{subfigure}{0.33\textwidth}
        \caption*{\large Isolated ($t = 3.91~\mathrm{Gyr}$)}
        \includegraphics[width=\textwidth]{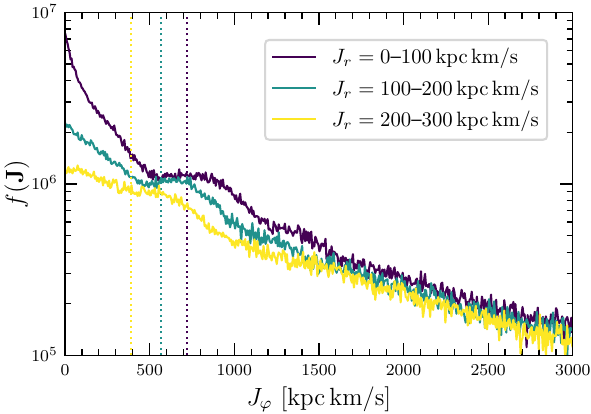}
      \end{subfigure}
      \begin{subfigure}{0.33\textwidth}
        \caption*{\large p00 ($t = 3.91~\mathrm{Gyr}$)}
        \includegraphics[width=\textwidth]{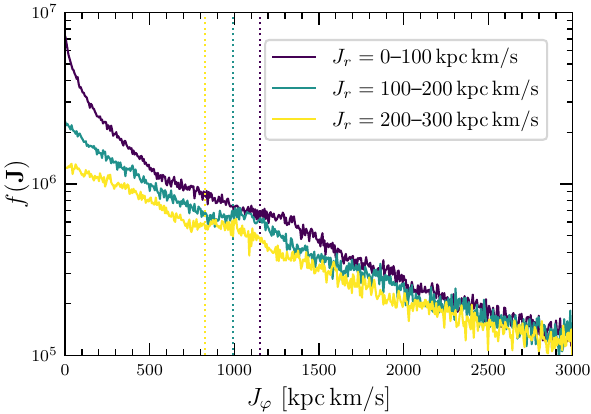}
      \end{subfigure}
    \\[1ex]
      \begin{subfigure}{0.33\textwidth}
        \includegraphics[width=\textwidth]{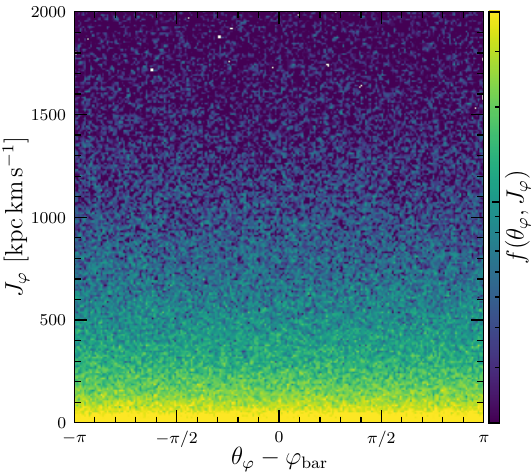}
      \end{subfigure}
      \begin{subfigure}{0.33\textwidth}
        \includegraphics[width=\textwidth]{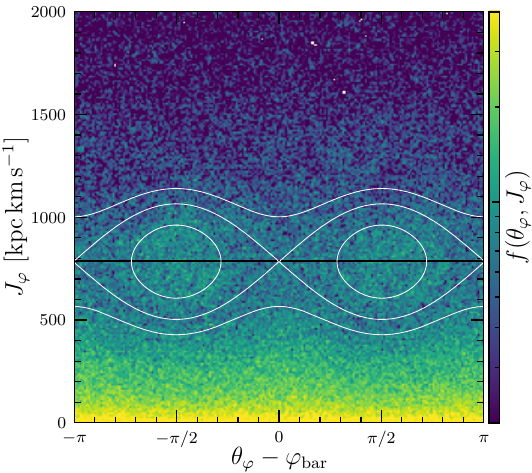}
      \end{subfigure}
      \begin{subfigure}{0.33\textwidth}
        \includegraphics[width=\textwidth]{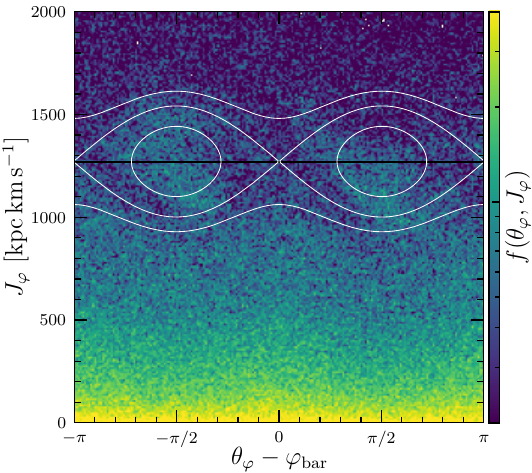}
      \end{subfigure}
      \vspace{2mm}
    \caption{Top panels: Distribution of halo particles in azimuthal action $J_\varphi$ ($z$-angular momentum) for the initial condition (left), the isolated model (middle), and the perturbed model p00 (right) at $t=3.91 \,\mathrm{Gyr}$. We select particles with $J_z < 100 \,\mathrm{kpc}\,\mathrm{km}\,\mathrm{s}^{-1}$ and plot the distribution for three different ranges in $J_r$ as denoted in the figure. The dotted vertical lines mark the position of the bar's corotation resonance. The initial distribution declines monotonically towards large $J_\varphi$, while the distribution in the isolated model is almost flat near the resonance, indicating the nonlinear saturation of dynamical friction. The gradient is recovered in the perturbed model. Bottom panels: The distribution in azimuthal angle-action space for $J_r < 100 \,\mathrm{kpc}\,\mathrm{km}\,\mathrm{s}^{-1}$. The white curves mark the contours of the Hamiltonian in the pendulum approximation (Appendix \ref{sec:pendulum_Hamiltonian}). The resonant distribution is phase mixed (saturated) in the isolated model, but remains unmixed in the perturbed model. {\textbf{Alt text}: Three line plots in the three top panels. The three lines indicate the values for three different ranges of $J_r$, $<100$, 100--200, and 200--300 kiloparsec kilometer per second. The x-axis is for $J_{\phi}$, ranging from 0 to 3000 kiloparsec kilometer per second. The y-axis is the fraction, $f(J)$, ranging between $10^5$ and $10^7$. Three density maps of distribution function, $f(\theta_{\phi}, J_{\phi})$, in the bottom panels. The $x$-axis is for the angle, $\theta _\varphi - \varphi_{\rm bar}$, from negative $\pi$ to positive $\pi$. The $y$-axis shows $J_{\phi}$, ranging from 0 to 2000 kiloparsec kilometer per second.}}
    \label{fig:phase_space_distribution}
\end{figure*}

As described in the Introduction (Section \ref{sec:introduction}), the temporal pause of the bar evolution in the isolated model suggests that dynamical friction has saturated and that the system has entered a metastable state. In contrast, the perturbed models appear to have avoided this by being continuously disturbed. To verify this, we looked into the phase-space distribution of the dark matter halo near the corotation resonance, which provides the largest torque among all resonances \citep{Athanassoula2003,Chiba2022Oscillating}. The top panels of Figure\,\ref{fig:phase_space_distribution} show the distribution of halo particles in azimuthal action $J_\varphi$ ($z$-angular momentum). We transformed coordinates using the St\"ackel fudge \citep{Binney2012Stackel} as implemented in \textsc{AGAMA} \citep{agama} and selected particles with relatively low inclination (vertical action $J_z < 100 \,\mathrm{kpc}\,\mathrm{km}\,\mathrm{s}^{-1}$) and plotted the distribution for three different ranges in radial action $J_r$. The left plot shows the initial distribution, which has a monotonic decline towards large $J_\varphi$. The middle panel shows the distribution in the isolated model at $t=3.91 \,\mathrm{Gyr}$. Evidently, the distribution near the resonance, marked by the dotted vertical lines, is flat, confirming the nonlinear saturation. Note that the resonance shifts towards small $J_\varphi$ as $J_r$ is increased, since $\Omega_\varphi$ declines with increasing $J_r$. The right plot shows the distribution in the perturbed model p00 at the same time $t=3.91 \,\mathrm{Gyr}$. The gradient was slightly recovered in the perturbed model, implying that friction is operating, consistent with the bar's behavior (Fig.\,\ref{fig:pattern_speed}).

The bottom row of Figure\,\ref{fig:phase_space_distribution} further shows the distribution of halo particles in the azimuthal angle-action space, where the azimuthal angle variable is measured with respect to the bar's angle. We selected particles with $J_r < 100 \,\mathrm{kpc}\,\mathrm{km}\,\mathrm{s}^{-1}$ and $J_z < 100 \,\mathrm{kpc}\,\mathrm{km}\,\mathrm{s}^{-1}$. 
To illustrate the nonlinear dynamics occurring in this phase space, we overlaid the contours of the Hamiltonian using the pendulum approximation in white (see Appendix~\ref{sec:pendulum_Hamiltonian} for details). In the absence of time-dependent perturbations other than the bar's rigid rotation, stars are expected to approximately follow these contours. As in the case of a simple pendulum, the phase space is divided into regions of libration, where trapped stars oscillate about the resonance center, and circulation, where untrapped stars far from the resonance circulate freely.
In the isolated model (middle bottom), we see two blobs of trapped particles near the resonance (black line), clustered around the bar's minor axis ($\pm \pi/2$). Within the limits of our particle resolution, the distribution appears well mixed along the motion of libration. The distribution is asymmetric with respect to the bar's major axis, meaning that the net torque on the bar is zero, as expected in a metastable state. In contrast, in the perturbed model (right bottom), the distribution near the resonance is distorted and exhibits certain asymmetries around the bar.

\begin{figure*}
    \centering
    \includegraphics[width=\textwidth]{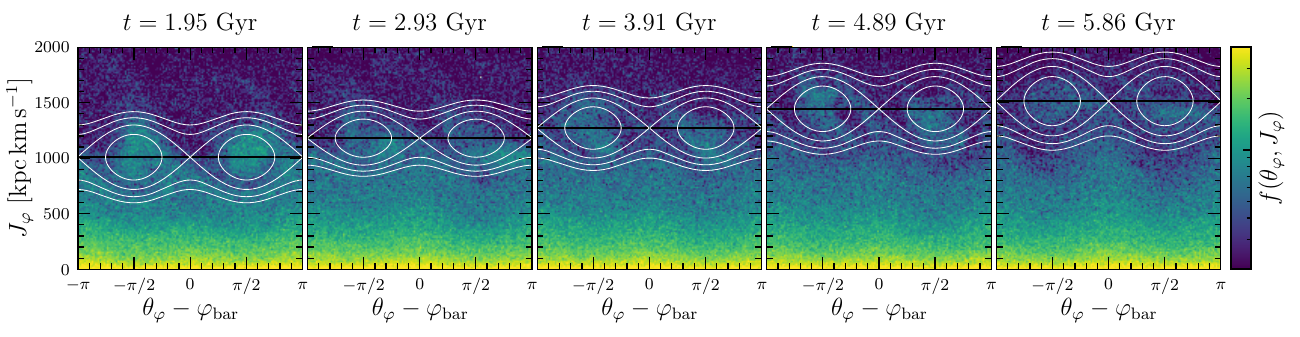}
      \vspace{2mm}
    \caption{Time evolution of the halo's azimuthal angle-action distribution in the perturbed model p00. The black lines mark the position of the bar's corotation resonance, while the white curves show the contours of the pendulum Hamiltonian. The distribution near the resonance never fully phase-mixes, resulting in sustained dynamical friction. {\textbf{Alt text}: Same as the bottom panels in Fig.~\ref{fig:phase_space_distribution}. Five panels show the distributions for t = 1.95, 2.93, 3.91, 4.89, and 5.86 Gyr.}}
    \label{fig:phase_space_distribution_time_evolution}
\end{figure*}

Figure\,\ref{fig:phase_space_distribution_time_evolution} further examines the time evolution of the halo's distribution of the perturbed model p00. The distribution exhibits a phase spiral around the resonance similar to that predicted analytically by \cite{Chiba2022Oscillating}. However, the phase spiral does not seem to tighten and approach a phase-mixed state as seen in the isolated model. This implies that the resonance is being persistently perturbed even after the dwarf has moved far away. We probe this further in the next section.

\subsection{Fluctuations in the dark matter halo induced by the dwarf fly-by}
\label{sec:halowake}

\begin{figure*}
    \centering
    \includegraphics[width=\textwidth]{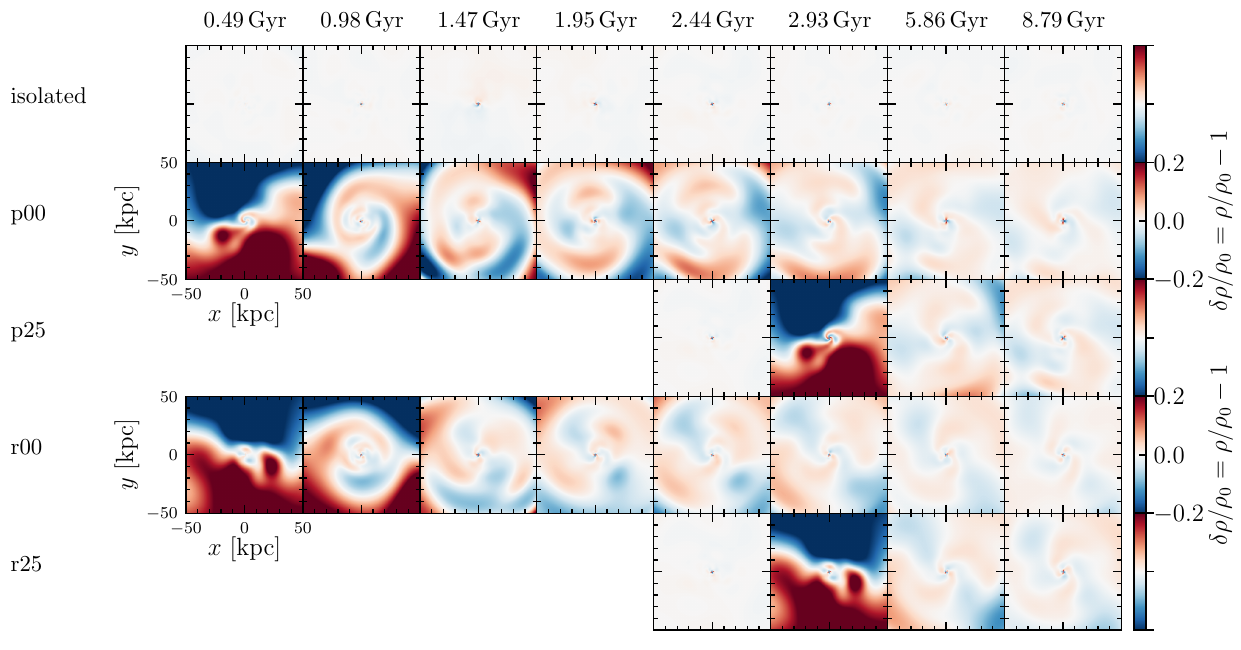}
    \caption{Fractional density perturbation in the dark matter halo at the galactic mid-plane $z=0$. A significant dipole perturbation is seen just after the dwarf encounter. After the encounter, the halo exhibits a persistent quadrupole perturbation that is very slowly rotating. {\textbf{Alt text}: Five panels in each column for models isolated, p00, p25, r00, and r25 models. Eight panels in each row for 0.49, 0.98, 1.47, 1.95, 2.44, 2.93, 5.86, and 8.79 Gyr. The plot range is -50 to 50 kpc for the x- and y-axes. Almost no fluctuation in the isolated model, but the fluctuation remains until 8.79 Gyr in the perturbed models.}}
    \label{fig:halo_density}
\end{figure*}

\begin{figure*}
    \centering
    \includegraphics[width=\textwidth]{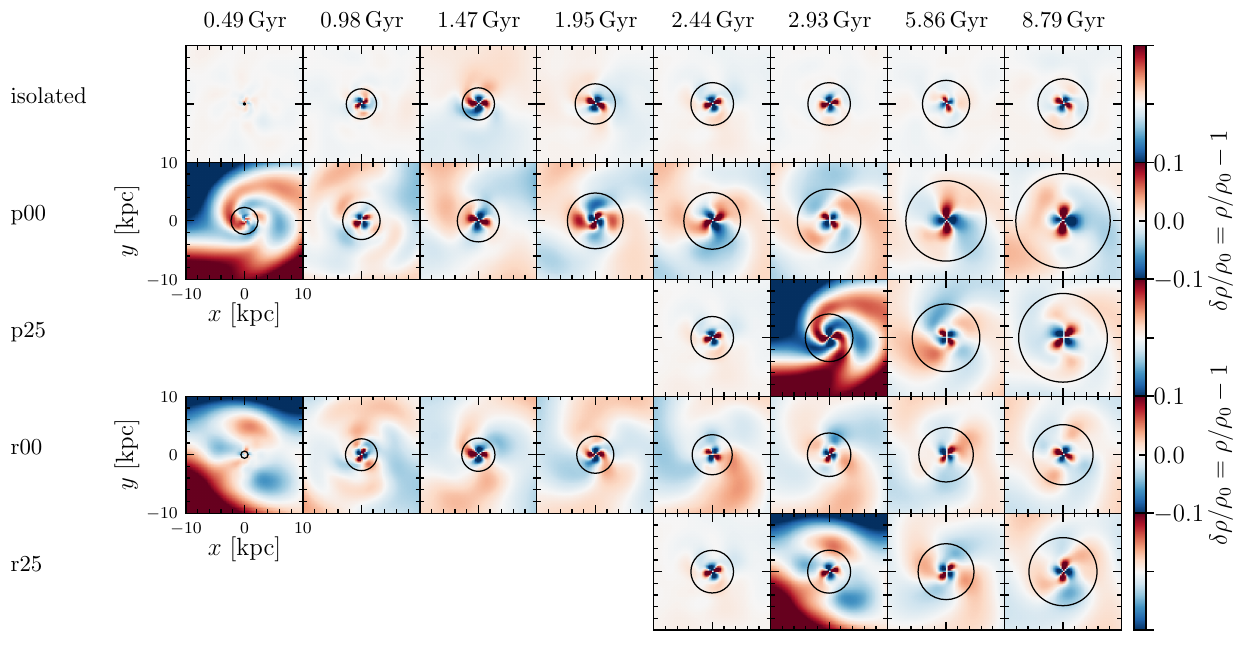}
    \caption{As in Fig.~\ref{fig:halo_density}, but zoomed into the inner disk region. Black circles mark the bar's corotation radius. {\textbf{Alt text}: Five panels in each column for models isolated, p00, p25, r00, and r25 models. Eight panels in each row for 0.49, 0.98, 1.47, 1.95, 2.44, 2.93, 5.86, and 8.79 Gyr. The plot range is -50 to 50 kpc for the x- and y-axes. Almost no fluctuation in the isolated model, but the fluctuation remains until 8.79 Gyr in the perturbed models.}}
    \label{fig:halo_density_zoomed}
\end{figure*}

Figure\,\ref{fig:phase_space_distribution_time_evolution} implies that the perturbations excited by the dwarf galaxy remain in the halo long after the impact time. To illustrate this, we present the fractional density perturbation in the halo in the galactic mid-plane ($z=0$) in Figures\,\ref{fig:halo_density} and \ref{fig:halo_density_zoomed}. The halo center is defined at the peak of the halo density at the inner core. The density is computed using a multipole expansion, and the fluctuation $\delta \rho = \rho - \rho_0$ is measured relative to the azimuthally averaged component $\rho_0$ at each snapshot.

Right after the dwarf impact, we see a spherical wake of the dwarf, along with a significant dipole perturbation resulting from the reflex motion of the inner halo, as seen in previous simulations of halo-satellite interactions (e.g., \cite{GaravitoCamargo2019Hunting}). As the dwarf flies away, the dipole perturbation gradually fades, and instead a quadrupole perturbation emerges. We find that the perturbation extends all the way to the halo center, allowing it to disturb particles near the bar's corotation resonances, indicated by black circles.
The very slow decay of the quadrupole perturbation implies that it is not a transient kinematic perturbation, but rather a self-sustained mode that is weakly damped \citep{Weinberg1989SelfGravitating}. 
Interestingly, we find that this quadrupole mode hardly rotates. We do not investigate the nature of this mode in detail here, but this is consistent with the recent study by \cite{Weinberg2023New}, which identified a quadrupole mode with zero pattern speed in NFW-like dark matter halos.
As suspected, the perturbation persists until the end of the simulation. 

\begin{figure*}
    \centering
    \includegraphics[width=0.33\textwidth]{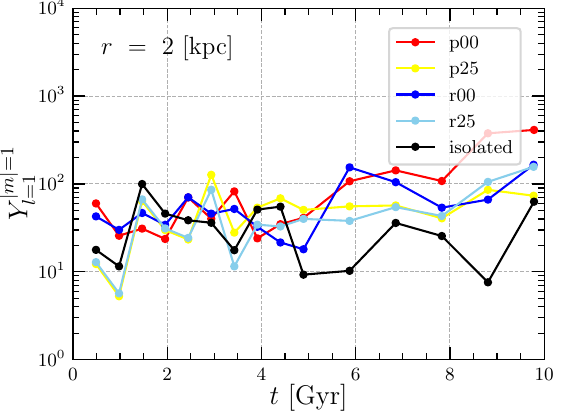}
    \includegraphics[width=0.33\textwidth]{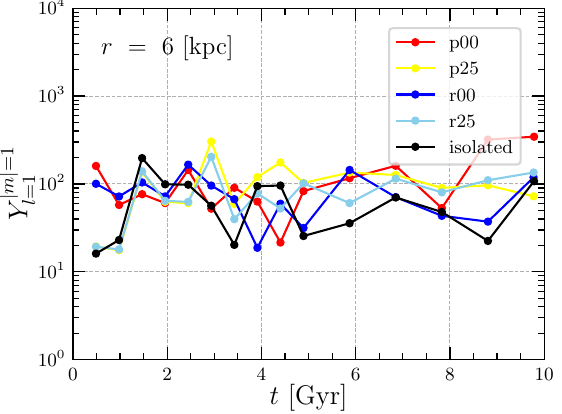}
    \includegraphics[width=0.33\textwidth]{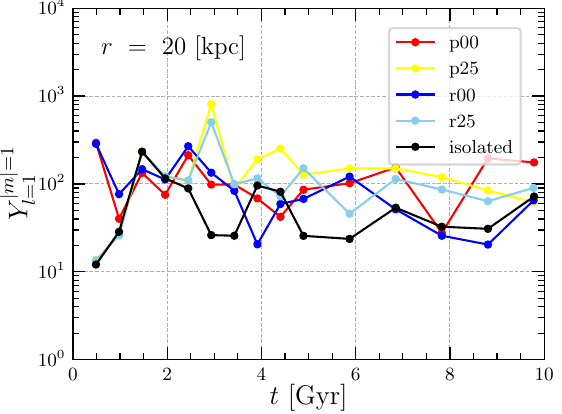}
    \includegraphics[width=0.33\textwidth]{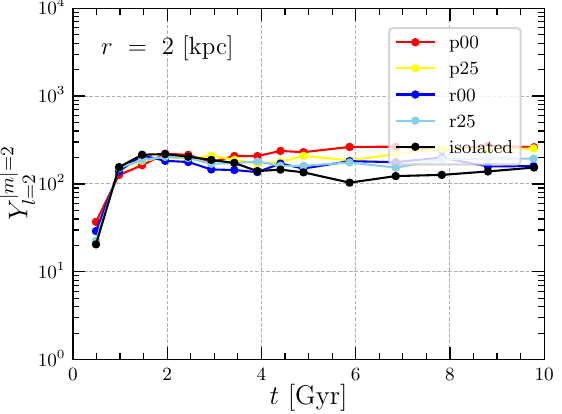}
    \includegraphics[width=0.33\textwidth]{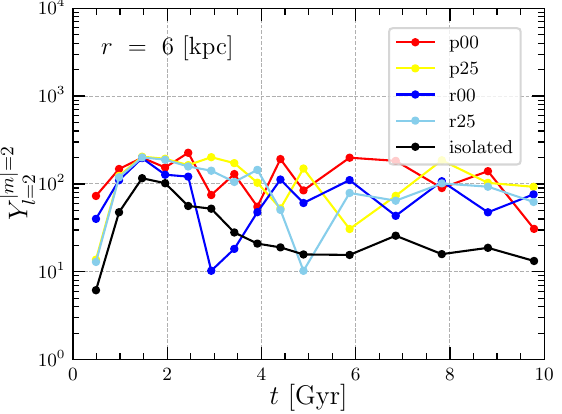}
    \includegraphics[width=0.33\textwidth]{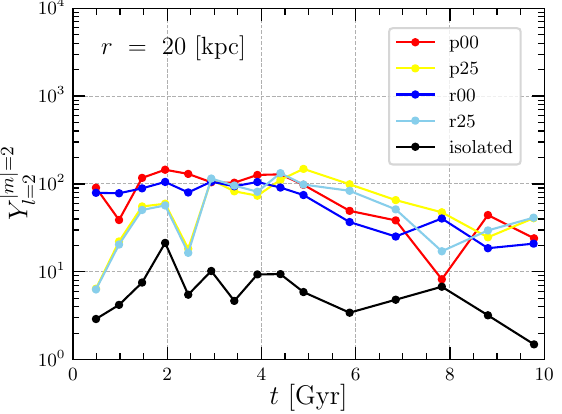}
    \vspace{2mm}
    \caption{Coefficient of spherical harmonics $Y_{l=1}^{|m|=1}$ (top panels) and $Y_{l=2}^{|m|=2}$ (bottom panels) at three different radii, increasing from left to right. $Y_l^{|m|}$ denotes the root-mean-square of $Y_l^m$ with $\pm m$. 
    \textbf{Alt text}: Time evolution of the spherical-harmonic amplitude $Y_l^{|m|}$ at several radii (panels from left to right: increasing radius). The top row shows the $(l,|m|)=(1,1)$ mode and the bottom row shows the $(l,|m|)=(2,2)$ mode.
    }
    \label{fig:Ylm}
\end{figure*}

Finally, to quantify the level of perturbation in the inner halo, we show in Figure\,\ref{fig:Ylm} the time evolution of the fluctuations in the halo potential, projected onto spherical harmonics at three different radii: $r=2$, 6, and 20 kpc from left to right. We show the coefficient $Y_l^{|m|}$ of the two dominant modes $(l,|m|)=(1,1)$ (top panels) and $(2,2)$ (bottom panels), where $Y_l^{|m|}$ denotes the root-mean-square of the $\pm m$ components. In the bar region ($r=2$ kpc; left panel), all models exhibit strong persistent fluctuations in the quadrupole mode ($|m|=2$) due to the formation of a dark bar, as seen in Figure \ref{fig:halo_density_zoomed}. In contrast, just outside the bar region near the corotation radius ($r=6$ kpc; middle panel), the quadrupole decays with time in the isolated model, whereas it persists in the perturbed models, consistent with the global quadrupole mode identified in the density maps (Figures \ref{fig:halo_density_zoomed} and \ref{fig:halo_density}). As expected, the amplitude of this global quadrupole fluctuation remains substantial even at larger radii ($r = 20$ kpc; right panel). At this radius, the fluctuations also show a peak in the dipole mode around the impact time.

\section{Summary}
\label{sec:summary}

Bar evolution in disk galaxies is governed by resonant coupling between the bar and the live dark matter halo, which generates dynamical friction and, as a result, slows the bar. Dwarf-galaxy fly-bys are common and naturally perturb both the disk and the dark matter halo, potentially modulating this coupling. To understand the reaction of the bar to the perturbation caused by a fly-by dwarf galaxy, we carried out fully self-consistent $N$-body simulations of a Milky Way-like disk galaxy with a live dark matter halo, comparing an isolated reference run to cases with a single coplanar fly-by of a self-gravitating dwarf galaxy on either a prograde or retrograde orbit. Pericenter passages were set before, during, and after the bar formation in the isolated model.
Our simulation showed: 
(i) For our bar-unstable galaxy model, fly-bys have little effect on the timing of bar formation; the bar formation essentially occurs on a timescale similar to that of the isolated model. 
(ii) Bars in perturbed models become slower than those in isolation. 
(iii) The final pattern speed is insensitive to encounter time but depends on orbital sense; prograde encounters produce a larger slowdown than retrograde ones. The difference in the orbits is consistent with idealized studies showing that prograde fly-bys exert stronger tidal forcing and more efficient resonant coupling than retrograde ones (e.g., \cite{1972ApJ...178..623T,1990A&A...230...37G,Lang2014}).
(iv) Angular-momentum budgets show complementary disk losses and halo gains, as expected for bar-halo resonant coupling (e.g., \cite{Athanassoula2003}).

To interpret these trends, we analyzed the phase-space evolution using angle-action variables. In the isolated run, the bar entered a saturated (metastable), weak-torque state once phase mixing flattens the distribution near the principal resonances, especially around corotation, and the slowdown stalled \citep{Sellwood2006Metastability,Chiba2022Oscillating,Banik2022Nonperturbative}. 
By contrast, a single dwarf passage excited long-lived perturbations in the halo's density and potential, which disturbed the halo particles near the bar resonances for long after the dwarf passage. As a result, the gradient of the distribution function near the bar resonances never fully flattened out. This kept the bar-halo friction active, thereby sustaining the bar's secular slowdown and growth. This picture is consistent with earlier simulations \citep{Sellwood2006Metastability} and theoretical works \citep{Hamilton2023BarResonanceWithDiffusion}, which find that modest external disturbances can prevent friction from saturating.

The fragility of the bar's metastable state against satellite impacts was demonstrated by \cite{Sellwood2006Metastability}. In this paper, we showed that a satellite impact even \textit{prior} to bar formation can induce perturbations in the halo that can later prevent the bar from entering a metastable state. This finding may have important implications for the Milky Way: several lines of evidence place the bar's formation at about 8 Gyr ago \citep{Sanders2024epoch}, close in time to the Gaia-Sausage/Enceladus (GSE) infall at about 9--10 Gyr ago \citep{Belokurov2020biggestsplash}, suggesting that an early encounter could have conditioned the subsequent friction state of the bar-halo system. Cosmological simulations likewise indicated that GSE-like mergers with modest mass ratios can promote bar formation on an early timetable \citep{2024MNRAS.531.1520M}. Future work that combines chemo-dynamical constraints from Gaia astrometry and large-scale ground-based spectroscopy with controlled live-halo simulations spanning mass ratio, inclination, and repeated passages will test whether this resonance-gradient restoration operates in the Milky Way.

\begin{ack}
We thank the anonymous referee for constructive comments and suggestions that helped improve the clarity and presentation of this paper.
This research used computational resources of Pegasus provided by Multidisciplinary Cooperative Research Program in Center for Computational Sciences, University of Tsukuba.
\end{ack}

\section*{Funding}
This work was supported by JSPS KAKENHI Grant Nos. 22KJ0829, 22J11943, 24K07095, 25H00664 and 25H00394.
We acknowledge the grants PID2021-125451NA-I00 and CNS2022-135232 funded by
MICIU/AEI/10.13039/501100011033 and by ``ERDF A way of making Europe’’, by the ``European Union'' and by the ``European Union Next Generation EU/PRTR''.

\section*{Data availability} 
 The simulation snapshots are available upon request.

\appendix

\section{Pendulum Hamiltonian}
\label{sec:pendulum_Hamiltonian}

\newcommand {\Gyr}{\,{\rm Gyr}}
\newcommand {\kpc}{\,{\rm kpc}}
\newcommand {\kpckpcGyr}{\,{\rm kpc}^2\,{\rm Gyr}^{-1}}

\newcommand {\vJ}{{\bm J}}
\newcommand {\Jr}{J_r}
\newcommand {\JR}{J_R}
\newcommand {\Jz}{J_z}
\newcommand {\dJr}{\dot{J}_r}
\newcommand {\dJR}{\dot{J}_R}
\newcommand {\Jphi}{J_\varphi}
\newcommand {\dJphi}{\dot{J}_\varphi}
\newcommand {\Jpsi}{J_\psi}
\newcommand {\Jf}{J_{\rm f}}
\newcommand {\vJf}{{\bm J}_{\rm f}}
\newcommand {\Jfo}{J_{{\rm f}_1}}
\newcommand {\Jft}{J_{{\rm f}_2}}
\newcommand {\Js}{J_{\rm s}}
\newcommand {\dJf}{\dot{J}_\mathrm{f}}
\newcommand {\dJs}{\dot{J}_\mathrm{s}}
\newcommand {\vJres}{{\bm J}_{\rm res}}
\newcommand {\Jzres}{J_{z,{\rm res}}}
\newcommand {\Jsres}{J_{\rm s, res}}
\newcommand {\Jsreso}{J_{\rm s, res0}}
\newcommand {\Jsresi}{J_{\rm s, res1}}
\newcommand {\Jssep}{J_{\rm s, sep}}
\newcommand {\dJsres}{\dot{J}_{\rm s, res}}
\newcommand {\Jphires}{J_{\varphi, {\rm res}}}
\newcommand {\DJs}{\Delta J_{\rm s}}
\newcommand {\DJsmax}{\Delta J_{\rm s, max}}
\newcommand {\Jsmax}{J_{\rm s, max}}
\newcommand {\dJres}{\delta J_{\rm res}}
\newcommand {\ddJres}{\delta \dot{J}_{\rm res}}

\newcommand {\vtheta}{{\bm \theta}}
\newcommand {\hvtheta}{\widehat{\bm \theta}}
\newcommand {\thetar}{\theta_r}
\newcommand {\thetaR}{\theta_R}
\newcommand {\thetaz}{\theta_z}
\newcommand {\thetaphi}{\theta_\varphi}
\newcommand {\thetapphi}{\theta'_\varphi}
\newcommand {\thetapsi}{\theta_\psi}
\newcommand {\thetas}{\theta_{\rm s}}
\newcommand {\dthetas}{\dot{\theta}_{\rm s}}
\newcommand {\ddthetas}{\ddot{\theta}_{\rm s}}
\newcommand {\thetaf}{\theta_{\rm f}}
\newcommand {\thetafo}{\theta_{{\rm f}_1}}
\newcommand {\thetaft}{\theta_{{\rm f}_2}}
\newcommand {\dtheta}{\dot{\theta}}
\newcommand {\ddtheta}{\ddot{\theta}}
\newcommand {\thetares}{\theta_{\rm res}}
\newcommand {\thetasres}{\theta_{\rm s,res}}
\newcommand {\thetasep}{\theta_{\rm sep}}
\newcommand {\thetasepm}{\theta_{\rm sep}^{-}}
\newcommand {\thetasepp}{\theta_{\rm sep}^{+}}
\newcommand {\thetassep}{\theta_{\rm s,sep}}
\newcommand {\thetal}{\theta_\ell}
\newcommand {\thetac}{\theta_{\rm c}}
\newcommand {\thetalc}{\theta_{\ell/{\rm c}}}
\newcommand {\thetap}{\theta_{\rm p}}
\newcommand {\vthetaf}{{\bm \theta}_{\rm f}}
\newcommand {\dphidt}{\dot{\phi}}
\newcommand {\dphidtsep}{\dot{\phi}_{\rm sep}}

\newcommand {\Omegap}{\Omega_{\rm p}}
\newcommand {\Omegapo}{\Omega_{\rm p0}}
\newcommand {\Omegapi}{\Omega_{\rm p1}}
\newcommand {\dOmegap}{\dot{\Omega}_{\rm p}}
\newcommand {\ddOmegap}{\ddot{\Omega}_{\rm p}}
\newcommand {\OmegaR}{\Omega_R}
\newcommand {\Omegar}{\Omega_r}
\newcommand {\Omegaz}{\Omega_z}
\newcommand {\Omegazsep}{\Omega_{z,{\rm sep}}}
\newcommand {\Omegaphi}{\Omega_\varphi}
\newcommand {\Omegapsi}{\Omega_\psi}
\newcommand {\vOmega}{\mathbf \Omega}
\newcommand {\vOmegaf}{{\bm \Omega}_{\rm f}}
\newcommand {\Omegaf}{\Omega_{\rm f}}
\newcommand {\Omegafo}{\Omega_{{\rm f}_1}}
\newcommand {\Omegaft}{\Omega_{{\rm f}_2}}
\newcommand {\Omegas}{\Omega_{\rm s}}
\newcommand {\Omegassep}{\Omega_{\rm s,sep}}
\newcommand {\Omegal}{\Omega_\ell}
\newcommand {\Omegac}{\Omega_{\rm c}}
\newcommand {\Omegalc}{\Omega_{\ell/{\rm c}}}
\newcommand {\OmegaILR}{\Omega_{\rm ILR}}
\newcommand {\Omegares}{\Omega_{\rm res}}
\newcommand {\Omegazres}{\Omega_{z,{\rm res}}}

\newcommand {\phib}{\varphi_{\rm bar}}
\newcommand {\drm}{\,\mathrm{d}}
\newcommand {\pd}{\partial}
\newcommand {\e}{\,\mathrm{e}}

\newcommand {\vn}{{\bm n}}
\newcommand {\vN}{{\bm N}}
\newcommand {\vk}{{\bm k}}
\newcommand {\vkf}{{\bm k}_{\rm f}}
\newcommand {\vm}{{\bm m}}
\newcommand {\nR}{n_R}
\newcommand {\nr}{n_r}
\newcommand {\npsi}{n_\psi}
\newcommand {\nphi}{n_\varphi}
\newcommand {\nz}{n_z}
\newcommand {\kr}{k_r}
\newcommand {\kpsi}{k_\psi}
\newcommand {\kphi}{k_\varphi}
\newcommand {\NR}{N_R}
\newcommand {\Nr}{N_r}
\newcommand {\Npsi}{N_\psi}
\newcommand {\Nphi}{N_\varphi}
\newcommand {\Nz}{N_z}
\newcommand {\ks}{k_{\rm s}}
\newcommand {\kf}{k_{\rm f}}
\newcommand {\kfo}{k_{{\rm f}_1}}
\newcommand {\kft}{k_{{\rm f}_2}}
\newcommand {\mf}{m_{\rm f}}
\newcommand {\mfo}{m_{{\rm f}_1}}
\newcommand {\mft}{m_{{\rm f}_2}}

This appendix describes the Hamiltonian used to draw the phase-space contours in Figures~\ref{fig:phase_space_distribution} and \ref{fig:phase_space_distribution_time_evolution}. The underlying theory has been described extensively in the literature, and we refer the readers to e.g., \cite{Binney2018,Chiba2021,Hamilton2024} for details. Here we summarize only the minimal set of equations used in this paper.

Near the bar's corotation resonance $(\Omegaphi - \Omegap = 0$), the azimuthal angle variable in the bar's rotation frame, $\thetapphi  \equiv \thetaphi - \phib$, evolves very slowly since $\dot{\theta}'_\varphi = \Omegaphi - \Omegap \approx 0$. This angle is therefore referred to as the \textit{slow} angle. More generally, for an arbitrary resonance satisfying $(\boldsymbol{N}\cdot\boldsymbol{\Omega} - N_\phi \Omega_{\rm p} = 0)$, where $\vN$ is a triplet of integers, the slow angle can be defined as 
\begin{align}
    \theta_{\rm s} \equiv \boldsymbol{N}\cdot\boldsymbol{\theta} - N_\varphi \varphi_{\rm bar}.
  \label{eq:thetas}
\end{align}
The remaining two angles, $(\theta_r, \theta_z)$, evolve on much shorter timescales and are collectively denoted as the \textit{fast} angles, $\boldsymbol{\theta}_{\rm f}$. The corresponding conjugate actions are
\begin{align}
(\Js,\Jfo,\Jft)=\left(\frac{\Jphi}{\Nphi},\Jr - \frac{\Nr}{\Nphi} \Jphi, \Jz - \frac{\Nz}{\Nphi} \Jphi\right).
  \label{eq:slow_fast_actions}
\end{align}
Because the rapid oscillations associated with $\vthetaf$ occur on timescales much shorter than the evolution of $\thetas$, the slow evolution of the system can be described by a Hamiltonian that is averaged over the fast angles. Since the averaged Hamiltonian no longer depends on $\vthetaf$, it follows from the Hamilton's equations that the conjugate fast actions are conserved. Thus the fast-angle averaging effectively reduces the six-dimensional phase-space dynamics to a two-dimensional dynamics in the slow angle-action space. Since we are only interested in the dynamics near the resonance, the Hamiltonian can be further Taylor-expanded about the resonant action $\Jsres$. The resulting Hamiltonian then takes the form of that of a simple pendulum:
\begin{align}
  H(\thetas,\Js) 
  = G (\Js - \Jsres)^2 - F \cos (\thetas - \thetasres),
  \label{eq:aveH}
\end{align}
where

\begin{align}
  G \equiv \frac{\pd^2 H_0}{\pd \Js^2}=\frac{\pd (\vN\cdot\vOmega)}{\pd \Js}, ~~~ {\rm and} ~~~ F \equiv - 2 |\hat{\Phi}_{(1,0,0)}|,
  \label{eq:GF}
\end{align}
with $\hat{\Phi}_\vn$ representing the Fourier coefficient of the perturbed potential. Both quantities $G$ and $F$ are evaluated at the resonance. In practice, we compute $G$ using finite difference with a step size of $\Delta \Js = 5 \kpckpcGyr$, while $F$ is obtained via a Fourier transform evaluated over $100\times100\times100$ grid points in angle space.

\bibliographystyle{aasjournal}
\bibliography{example}

\end{document}